\documentclass{desyproc}
\usepackage{amsmath}
\usepackage{amssymb}
\usepackage{wrapfig}

\begin{document}
\title{Using Graphics Processing Units to solve the classical $N$-Body Problem in Physics and Astrophysics}

\author{{\slshape Mario Spera$^{1,2}$}\\[1ex]
$^1$INAF-Osservatorio Astronomico di Padova, Vicolo dell'Osservatorio 5, I-35122, Padova, Italy; \\
$^2$Sapienza, Universit\'a di Roma, P.le A. Moro 5, I-00165 Rome, Italy;\\
\\
\texttt{mario.spera@oapd.inaf.it}\,\,\,\,\,\,\,\, \texttt{mario.spera@live.it}}

\contribID{35}

\confID{7534}  
\desyproc{DESY-PROC-2014-05}
\acronym{GPUHEP2014}
\doi

\maketitle

\begin{abstract}
Graphics Processing Units (GPUs) can speed up the numerical solution of various problems in astrophysics
including the dynamical evolution of stellar systems; the performance gain can be more than a factor
100 compared to using a Central Processing Unit only. In this work I describe some strategies to speed
up the classical $N$-body problem using GPUs. I show some features of the $N$-body code \texttt{HiGPUs} as template code. In
this context, I also give some hints on the parallel implementation of a regularization method and I 
introduce the code \texttt{HiGPUs-R}. Although the main application of this work concerns astrophysics,
some of the presented techniques are of general validity and can be applied to other branches of physics such
as electrodynamics and QCD.
\end{abstract}

\section{Introduction}
\label{sec:introduction}
The $N$-body problem is the study of the motion of $N$ point-like particles interacting through their mutual 
forces that can be expressed according to a specific physical law. In particular, if the reciprocal 
interaction is well approximated by the Newton's gravity law, we refer to the classical, gravitational 
$N$-body problem. The differential equations that describe the kinematics of the $N$-body system are

\begin{equation}
\label{eq:nbody_system}
\begin{cases}
\mathbf{\ddot{r}}_i &= -G\sum\limits_{\substack{j=1 \\ j\neq i}}^{N}\frac{m_j}{{r_{ij}}^{3}}\mathbf{r_{ij}}\\
\mathbf{r}_i\left(t_0\right) &= {\mathbf{r}}_{i,0}\hspace{50pt} i = 1,2,...,N\\
\mathbf{\dot{r}}_i\left(t_0\right) &= {\mathbf{\dot{r}}}_{i,0}
\end{cases}
\end{equation}
where $t$ is the time, $\mathbf{r}_i$, 
$\mathbf{\dot{r}}_i$ and $\mathbf{\ddot{r}}_i$ are the position, the velocity and the acceleration 
of the $i$-th particle, respectively, $G$ is the universal gravitational constant, $m_j$ indicates the mass 
of the particle $j$,
${\mathbf{r}}_{i,0}$ and ${\mathbf{\dot{r}}}_{i,0}$ represent the initial position and velocity and $r_{ij}$ 
is the mutual distance between particle $i$ and particle $j$. Although we 
know that the solution of the system of equations (\ref{eq:nbody_system}) exists and is unique, we do not 
have its explicit 
expression. Therefore, the best way to solve the $N$-body problem is numerical. The numerical 
solution of the system of equations (\ref{eq:nbody_system}) is considered a challenge despite the 
considerable 
advances in both software development and computing technologies; for instance, it is still not 
possible to study the dynamical evolution of stellar systems composed of more than $\sim10^6$ objects without 
the need of theoretical approximations. The numerical issues come mainly from two aspects:
\begin{itemize}
\item {\em ultraviolet divergence (UVd)}: close encounters between particles ($r_{ij}\rightarrow0$) produce a 
divergent mutual force 
($F_{ij}\rightarrow\infty$). The immediate consequence is that the numerical time step must be very small in 
order to follow the rapid changes of positions and velocities with sufficient accuracy, slowing down the  
integration.
\item {\em infrared divergence (IRd)}: to evaluate the acceleration of the $i$-th particle we need to take 
into 
account all 
the other $N-1$ contributions 
because the gravitational force never vanishes ($F_{ij}\rightarrow 0 \Leftrightarrow r_{ij} \rightarrow 
\infty$). This implies that the $N$-body problem has a computational complexity of $O\left(N^2\right)$.
\end{itemize}
To control the effects of the UVd and smooth the behavior of the force during close 
encounters, a parameter $\epsilon$ ({\em softening parameter}) is introduced in the 
gravitational potential. This leads to an approximate expression for the reciprocal attraction which is

\begin{equation}
\mathbf{F_{i}}=-G\frac{m_im_j}{\left({r_{ij}}^2 + \epsilon^2\right)^{\tfrac{3}{2}}}\mathbf{r_{ij}}.
\end{equation}
In this way, the UVd is artificially removed paying a loss of resolution at spatial scales 
comparable to $\epsilon$ and below. An alternative approach concerns the usage of a regularization method, 
that is a coordinate transformation that modifies the standard $N$-body Hamiltonian removing the singularity 
for $r_{ij}=0$. We briefly discuss this strategy in Sec. \ref{sec:regularization}. 

On the other hand, the issues that come from the IRd can be overcome using:

\begin{enumerate}
\item {\em approximation schemes}: the direct sum of inter-particle forces is replaced by another 
mathematical expression with lower computational complexity. To this category belongs, for instance, 
the {\em tree scheme}, originally introduced by Barnes and Hut, which is one of the most known approximation 
strategies~\cite{barnes1986};
\item {\em hardware acceleration}: it is also possible to use more efficient hardware to speed up the force 
calculation maintaining the $O\left(N^2\right)$ computational complexity of the problem.
\end{enumerate}

For what concerns hardware advances, Graphics Processing Units (GPUs) can act as computing 
accelerators of the evaluation of the mutual forces. This approach is extremely efficient because a GPU is a 
highly parallel device which can have up to $\sim$ 3,000 cores and run up to $\sim$ 80,000 virtual units (GPU 
threads) that can execute 
independent instructions at the same time. Since the evaluations of mutual 
distances can be executed independently, a GPU perfectly matches the structure of the 
$N$-body problem. Nowadays, the overwhelming majority of $N$-body simulations are carried out 
exploiting the GPU acceleration.

\section{The direct $N$-body code \texttt{HiGPUs}}
In this section I give an overview of the most common strategies adopted to numerically solve the 
$N$-body problem using a GPU. I describe the \texttt{HiGPUs} code as an example of GPU optimized $N$-body 
code~\cite{dolcetta2013a}. \texttt{HiGPUs}\footnote{ 
\texttt{http://astrowww.phys.uniroma1.it/dolcetta/HPCcodes/HiGPUs.html}} is a direct summation $N$-body code 
that implements a Hermite 6th order time integrator. It is written combining utilities of C and 
C++ and it uses CUDA (or OpenCL), MPI and OpenMP to exploit GPU workstations as well as GPU clusters. The main features of \texttt{HiGPUs} and of other GPU optimized $N$-body codes are the following:

\textbf{1. THE HERMITE ALGORITHM}: the Hermite time integrators (4th, 6th and 8th order) represent the state of the art of direct $N$-body 
simulations(~\cite{aarseth2003},~\cite{nitadori2008}). In particular, the 4th order scheme is, by far, the 
most widely used algorithm in this context 
being particularly efficient in terms of ratio between computing time and accuracy. The Hermite integrators 
are based 
on Taylor series of positions and velocities and their most important feature is that they have high accuracy 
even though they need to evaluate the distances between particles just once per time step. This is a 
huge advantage in using Hermite integrators if we consider that, for example, a standard Runge-Kutta method 
needs to evaluate accelerations 4 times per integration step and it is ``only'' 4th order accurate. 
\begin{wrapfigure}{r}{0.45\textwidth}
	
	\begin{center}
		\includegraphics[width=0.45\textwidth]{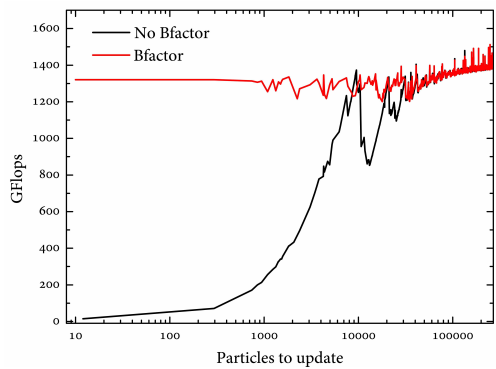}
	\end{center}
	
	\caption{The effects of the \texttt{Bfactor} optimization on the performance of a GeForce GTX TITAN Black 
		(measured in billions of floating point operations per second, GFlops) as a function of the number of 
		particles 
		that need to be updated (according to the block time steps distribution). The results were obtained using the 
		$N$-body code \texttt{HiGPUs} with $N\simeq 260,000$. }
	\label{fig:fig1}
\end{wrapfigure}
\textbf{2. BLOCK TIME STEPS}: stars in an $N$-body system can have very different accelerations; this corresponds to have a large variety of evolutionary time-scales. In this context, it 
is 
convenient to assign to all the objects their own time step which becomes a function of the physical 
parameters 
that describe the kinematic state of the corresponding particle. In order to avoid time 
synchronization issues among the $N$ bodies and to simplify the parallelization process, the time step is 
forced to be a power of two. Thus, particles are sub-divided in several groups 
(blocks) that 
share the same time step~\cite{aarseth2003}. In this way we need to update positions and velocities only of 
$m\leq N$ 
particles per time step; in particular, bodies with smaller time steps will be updated more often than 
particles with bigger time steps for which the kinematic state will be estimated using Taylor expansions 
only. This implies that the computational complexity per time step is reduced from $O\left(N^2\right)$ to 
$O\left(mN\right)$.

\textbf{3. THE \texttt{Bfactor} VARIABLE}: another important aspect concerns the GPU load. For example, a 
GeForce GTX TITAN Black GPU can run a maximum of 30,720 
threads in parallel, therefore we need to fittingly distribute the 
work load to fully exploit this kind of GPU. When using the block 
time steps strategy it is quite common to have $m<30,720$ therefore we introduce the variable 
\texttt{Bfactor} that can increase the number of running threads and further split 
and distribute the work load among the GPU 
cores. For instance, if we have $m<30,720$ and \texttt{Bfactor}$=B$ we run $mB$ threads and each thread 
calculates the accelerations due to $N/B$ bodies. The optimal \texttt{Bfactor} value must be determined step 
by step. We show in Fig. 
\ref{fig:fig1} the differences in terms of GPU performance running a typical $N$-body simulation 
with and without the \texttt{Bfactor} variable. It is evident that, for a GeForce GTX TITAN Black, when the 
particles that must be updated are $m\lesssim30,720$ we obtain significantly higher performance when the 
\texttt{Bfactor} optimization is turned on. A similar optimization strategy can be found in~\cite{nyland2007}.

\textbf{4. PRECISION}: it is well known that the maximum theoretical performance of all the GPUs in double 
precision (DP) is lower than 
their 
capability to execute single precision (SP) operations. For $N$-body problems it is important to 
use DP to calculate reciprocal distances and to 
cumulate accelerations in order to reduce round-off errors as much as possible. All 
the other instructions (square roots included) can be executed in SP to speed up the 
integration. Some authors use an alternative approach based on an emulated double precision arithmetic (also 
known as double-single precision or simply DS). In this strategy a DP variable is replaced 
with two, properly handled, SP values; in this way only SP quantities are used against a slightly larger 
number 
of operations that must be executed~\cite{gaburov2009}.

\textbf{5. SHARED MEMORY}: the GPU shared memory (SM) 
is a 
limited amount of memory (in general $\lesssim 65$kB) and it is shared between all the GPU threads in the same 
block and can be used for fast data transactions (on average, SM is about 10 times 
faster than ``standard'' memory). During the evaluation of the $N$-body accelerations, the best strategy is 
to 
cyclically load SM until all the pair-wise forces are computed.

\subsection{Performance results}
\begin{figure}
\begin{center}
\includegraphics[width=0.48\textwidth]{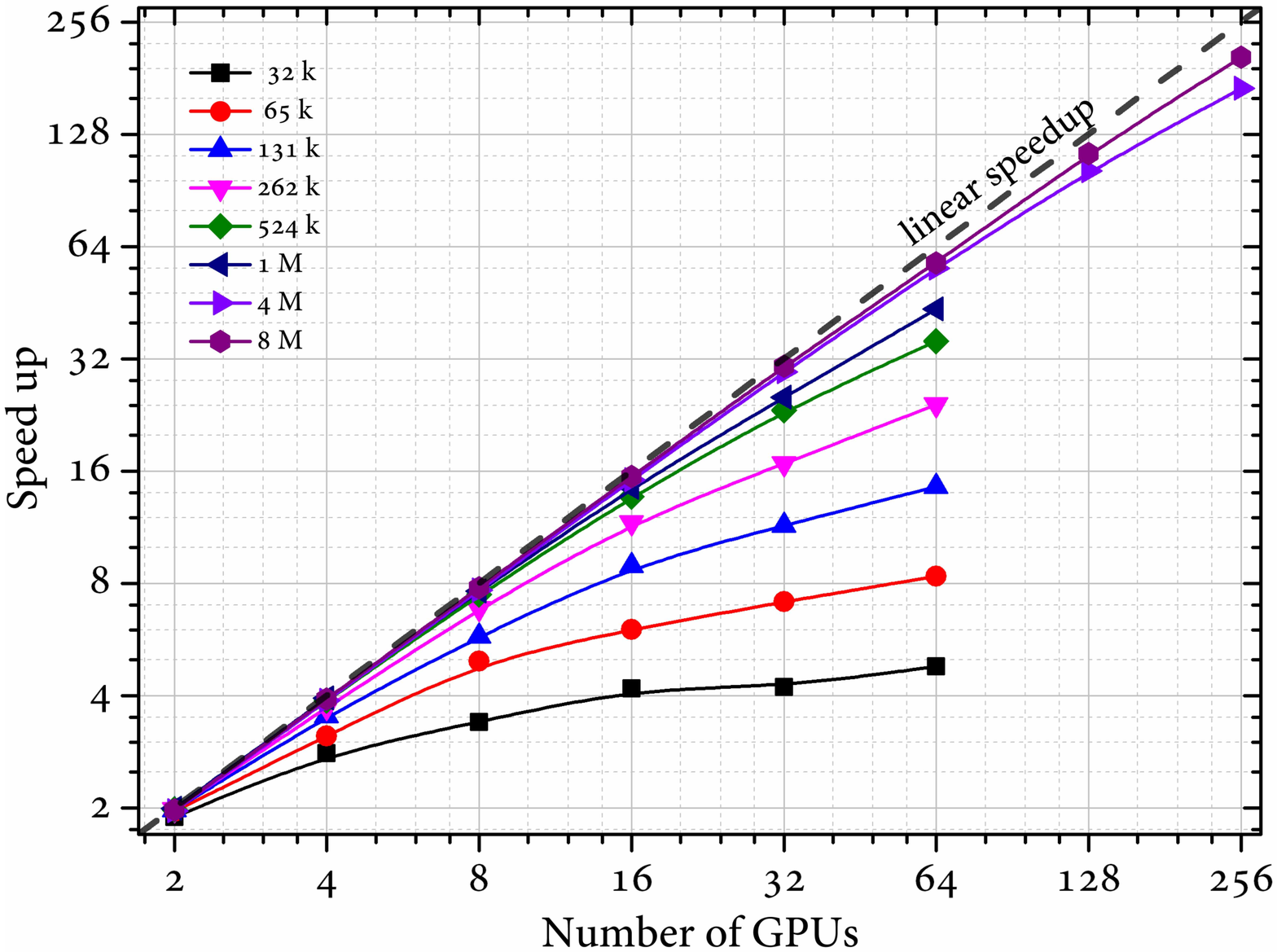}
\includegraphics[width=0.48\textwidth]{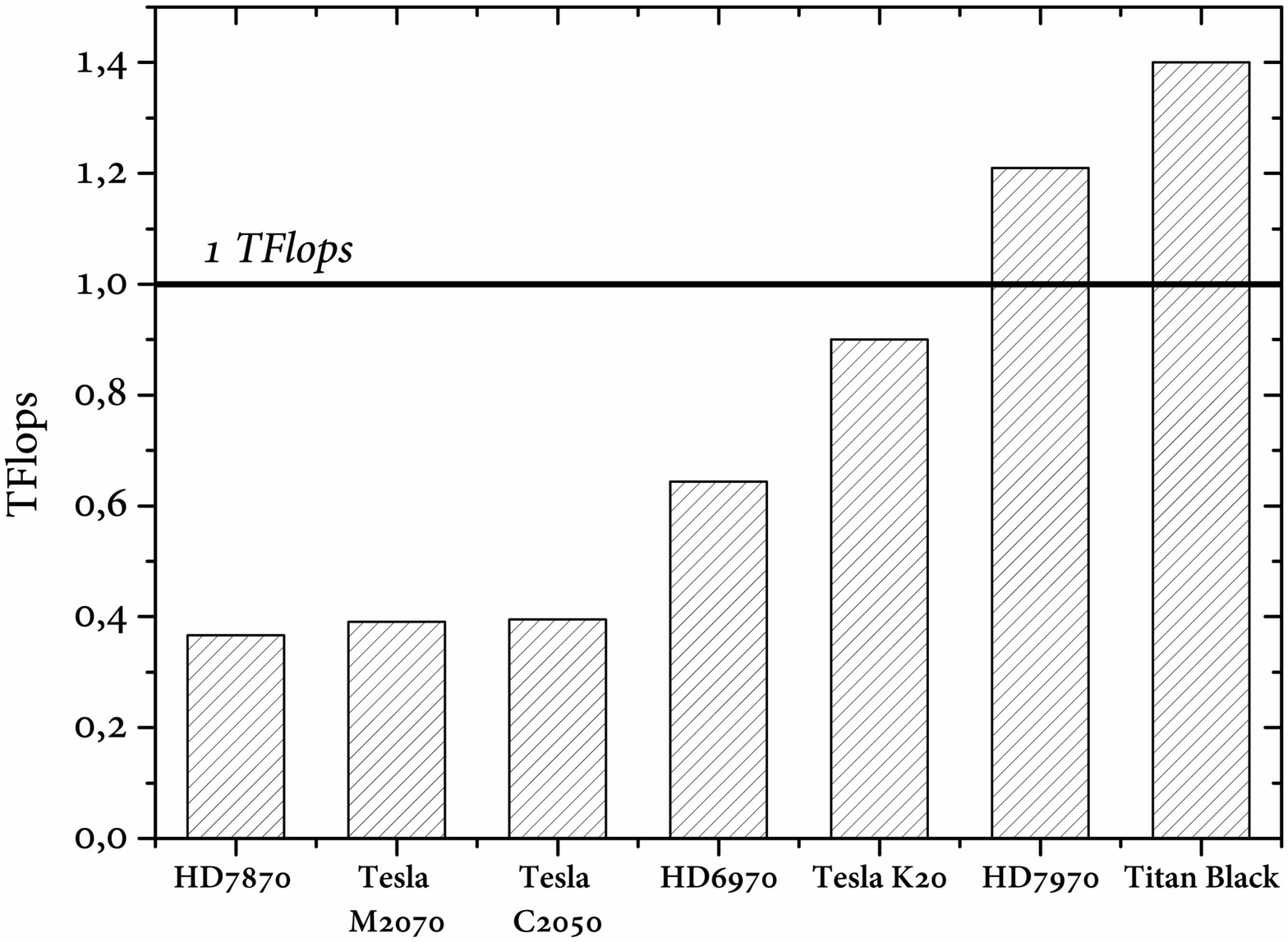}
\caption{Left panel: speed up of the code \texttt{HiGPUs} as a function of the number of used GPUs for 
different $N$-body systems with $3.2\times{10^4}\lesssim N \lesssim 8\times{10^6}$. The dashed line corresponds 
to the maximum computing efficiency (linear speedup). Right panel: performance (in TFlops) of \texttt{HiGPUs} 
running on single, different GPUs. For the scalability tests we used the IBM DataPlex DX360M3 Linux Infiniband Cluster provided by the Italian 
supercomputing consortium CINECA.}
\label{fig:fig2and3}
\end{center}
\end{figure}Fig. \ref{fig:fig2and3} shows the scalability of the code \texttt{HiGPUs} on a GPU cluster (left 
panel) and its performance using single, different GPUs (right panel). Form Fig. \ref{fig:fig2and3}, it is 
apparent that GPUs are extremely well suited to solve the $N$-body problem: we reach a computing efficiency 
of $\sim 92\%$ using 256 GPUs and $\sim 8$ million bodies and a sustained performance of $\sim 1.4$ TFlops on 
just one GPU (GeForce GTX TITAN Black). More details can be found in~\cite{dolcetta2013a} 
and~\cite{dolcetta2013b}.

\section{Regularization}
\label{sec:regularization}
Close encounters between (two or more) particles are critical in $N$-body simulations because of 
the UVd of the 
gravitational force. An attempt to remove the small-scale singularity of the interaction potential is 
referred as an attempt of {\em regularization}. The Burdet-Heggie method 
(~\cite{burdet1967},~\cite{heggie1973}), the Kustaanheimo-Stiefel algorithm~\cite{kusta1965} 
or the Mikkola's algorithmic formulation (MAR,~\cite{mikkola1999}) are some of the most famous examples of 
regularization. In 
general, all these 
methods are quite expensive in terms of implementation effort and computing time but, if we use them to 
integrate few bodies only, they become both faster and much more accurate than standard 
$N$-body integrators. It is not convenient to implement regularization methods on a GPU
because of their mathematical construction and because they can integrate, in general, a maximum of few tens 
of bodies. Nevertheless, during the dynamical evolution of an $N$-body system, we can identify the groups of 
particles that are in tight systems or that are experimenting a close encounter, and regularize 
them. \begin{figure}
\begin{center}
	\includegraphics[width=0.48\textwidth]{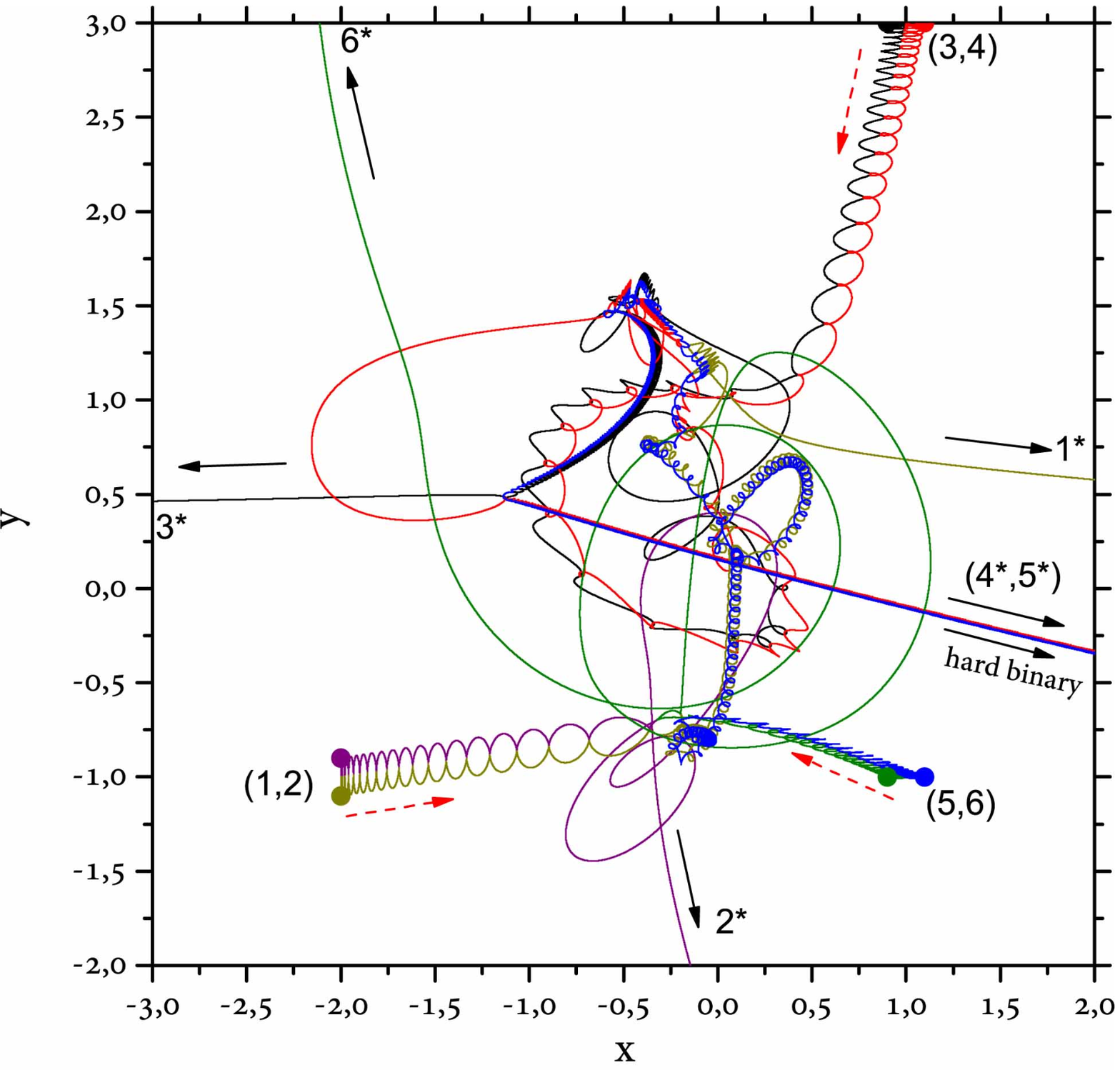}
	\includegraphics[width=0.48\textwidth]{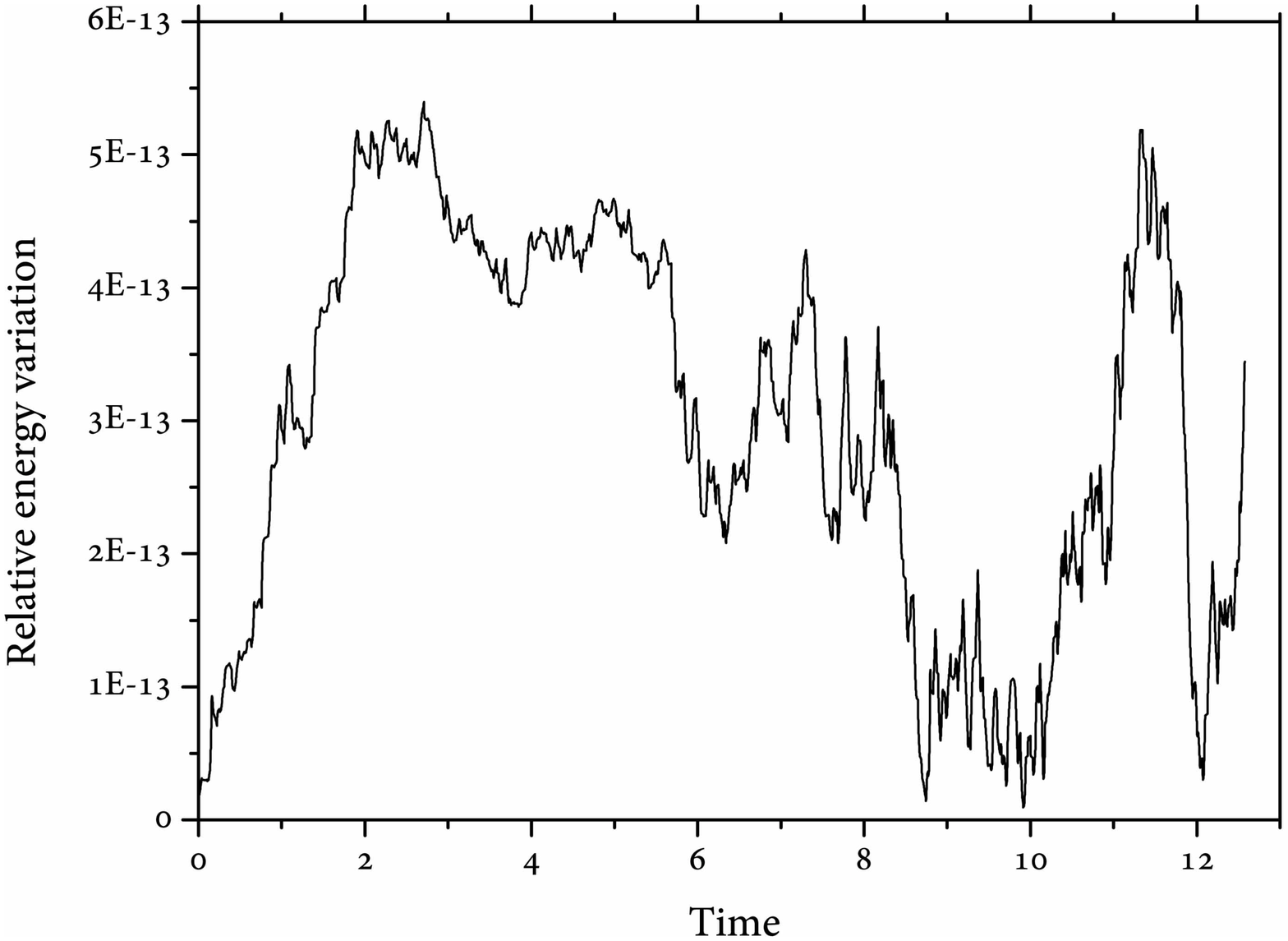}
	\caption{Left panel: trajectories obtained for the modified version of the Pythagorean 3-body problem, where the three particles are replaced with three binaries, using a regularized algorithm. 
		Three binary systems, indicated with (1,2), (3,4) and (5,6), are initially placed at the vertices of a right  
		triangle with null velocities. During the dynamical evolution, the particles 1, 2, 3 and 6 are ejected form 
		the system with high velocities (outside directions are indicated with arrows) while particles 4 and 5 form a 
		very hard binary. Right panel: relative energy variation during the integration of the Pythagorean 
		problem.}
	\label{fig:fig4and5}
\end{center}
\end{figure}This process can be done in parallel by the CPU, by means of OpenMP, establishing a 1 to 1 
correspondence between groups that must be regularized and CPU threads. At the same time, given that the GPU 
kernels are asynchronous, the regularization process can be performed while the GPU works in background. This 
describes the parallel scheme adopted to implement the MAR in the code \texttt{HiGPUs-R} which is still under 
development. A test application to demonstrate the advantages of regularization is shown in Fig. 
\ref{fig:fig4and5}. It represents a modified version of the so called Pythagorean 
3-body problem (e.g.~\cite{mikkola1993}) integrated with \texttt{HiGPUs-R}. Standard $N$-body integrators, 
such as the Hermite 4th order scheme, cannot evolve this system,
because either the time step becomes prohibitively small throughout the dynamical evolution, or, if we fix a minimum time step, the solution is completely inaccurate. The only 
chance is to use a regularized code which is very fast ($\sim 10$ seconds of simulations to 
obtain the trajectories in the left panel of Fig. \ref{fig:fig4and5}) and maintains a very good total energy 
conservation (see the right panel of Fig. \ref{fig:fig4and5}).

\section{Conclusions}
In this work I have presented and discussed the main strategies adopted to speed up the numerical solution of 
the $N$-body problem using GPUs. I have also shown the main advantages in using 
regularization methods and described a new parallel scheme to implement the Mikkola's algorithmic 
regularization in the context of a GPU $N$-body code. I have used the direct $N$-Body code \texttt{HiGPUs} as 
reference and I have given an overview of the code \texttt{HiGPUs-R} that is a new regularized version of 
\texttt{HiGPUs}. The development of fast and regularized $N$-body codes such 
as \texttt{HiGPUs-R} is of fundamental importance to investigate a large 
number of astrophysical problems (ranging from the dynamical evolution of star clusters to the formation of double black hole binaries).

\section{Acknowledgments}
MS thanks Michela Mapelli and Roberto Capuzzo Dolcetta for useful discussions, and acknowledges financial 
support from the MIUR through grant FIRB 2012 RBFR12PM1F.


\begin{thebibliography}{99}
\bibitem{barnes1986} J.~Barnes and P.~Hut, Nature {\bf 324} 446 (1986)
\bibitem{dolcetta2013a} R.~Capuzzo-Dolcetta, M.~Spera and D.~Punzo, JCP {\bf 236} 580 (2013)
\bibitem{nitadori2008} Nitadori, K., \& Makino, J., New Astronomy {\bf 13} 498 (2008)
\bibitem{aarseth2003} S.J.~Aarseth, {\it Gravitational N-body simulations: tools and algorithms}, Cambridge 
Univ. Press, UK (2003)
\bibitem{nyland2007} L.~Nyland, M.~Harris and J.~Prins, GPU gems 3 {\bf 31} 677 (2007)
\bibitem{gaburov2009} E.~Gaburov, S.~Harfst and S.~Portegies Zwart, New. Ast. {\bf 14} 630 (2009)
\bibitem{dolcetta2013b} R.~Capuzzo-Dolcetta, M.~Spera, Comp. Phys. Comm. {\bf 184} 2528 (2013)
\bibitem{burdet1967} C.A.~Burdet, Z. Angew. Math. Phys. {\bf 18} 434 (1967)
\bibitem{heggie1973} D.C.~Heggie, Astr. and Space Science Lib. {\bf 39} 34 (1973)
\bibitem{kusta1965} P.~Kustaanheimo and E.~Stiefel, Jour. f\"ur Reine und  Angew. Math. {\bf 218} 204 (1965)
\bibitem{mikkola1999} S.~Mikkola and K.~Tanikawa, Celest. Mech. Dyn. Astr. {\bf 74} 287 (1999)
\bibitem{mikkola1993} S.~Mikkola and S.J.~Aarseth, Celest. Mech. Dyn. Astr. {\bf 57} 439 (1993)
\end{thebibliography}
\end{document}